\documentclass[]{aastex631}
\usepackage{etoolbox}
\usepackage{amsmath}
\usepackage{CJK}
\AtBeginEnvironment{longtable}{\tiny}

\shorttitle{Possible Correlation between Metallicity and Color for Late-M/L Dwarfs}

\graphicspath{{./}{figures/}}

\begin{document}
\begin{CJK*}{UTF8}{gbsn}
\title{A Possible Correlation between Metallicity and Near-IR Color for Late-M and L Dwarfs}

\author[0000-0002-7555-4310]{Ruihan Zhang (张瑞涵)}
\affiliation{Institute for Astronomy, University of Hawai'i\\
Hilo, HI 96720, USA}

\author[0000-0003-2232-7664]{Michael C. Liu}
\affiliation{Institute for Astronomy, University of Hawai'i\\
Honolulu, HI 96822, USA}

\author[0000-0002-3726-4881]{Zhoujian Zhang (张周健)} \thanks{NASA Sagan Fellow}
\affiliation{Department of Astronomy \& Astrophysics, University of California, Santa Cruz, CA 95064, USA }

\correspondingauthor{Ruihan Zhang}
\email{rzhang9@hawaii.edu}

\begin{abstract}

We examine the relationship between metallicity and $J-K$ color for 64 benchmark late-M and L dwarfs, all of which are wide companions to higher mass stars, and 6 of which are new discoveries. We assess the correlation between the $\Delta(J-K)$ color anomaly (the difference of an object's $J-K$ color with the median color for field objects of the same spectral type) and the host star metallicity to investigate how metallicity affects ultracool photospheres. Using Spearman's rank correlation test and Student's t test, the late-M dwarf (L dwarf) sample's $\Delta(J-K)$ and metallicity show a positive correlation with 95\% (90\%) confidence level. A linear fit to color anomaly as a function of metallicity finds a slope of $0.17\pm0.07$ for the late-M dwarfs and a slope of $0.20^{+0.07}_{-0.08}$ for the L dwarfs. We also computed the $\Delta(J-K)$ versus metallicity relationship predicted by multi-metallicity model spectra generated using Drift-Phoenix. The modeled late-M dwarfs show a slope of 0.202$\pm$0.03, which is close to our observational results, but the modeled L dwarfs show a slope of 0.493$\pm$0.02, steeper than our observational results. Both our empirical results and the models indicate that more metal-rich objects should appear redder photometrically. We speculate that higher metallicity drives more condensate formation in these atmospheres, thus making these ultracool dwarfs appear redder.

\end{abstract}

\keywords{Brown dwarfs(185) --- L dwarfs(894) --- Metallicity(1031) --- Photometry(1234)}

\section{Introduction} \label{sec:intro}
Understanding the atmospheres of ultracool dwarfs (late-M, L, and T dwarfs) is particularly interesting because they span similar temperature and mass regimes as giant exoplanets. Thus learning more about ultracool dwarfs would help us better understand the atmospheres of the giant planets outside of our solar system. In contrast to higher mass stars, which follow a narrow sequence on the color-magnitude diagrams (CMD) in the near-IR (NIR), the photometry of ultracool dwarfs show a significant spread. Some of the spread can be attributed to variations in surface gravity (\citealt{Knapp2004}, \citealt{Liu2016}), which is related to the mass and age spread of these objects. Low gravity objects, or young objects, appear redder in NIR photometry compared to the field objects of the same spectral types. The viewing geometry of the ultracool dwarfs also plays a role in creating color anomalies in ultracool dwarf photometry (\citealt{Vos2017}, \citealt{Mateo2022}). When ultracool dwarfs are viewed at larger inclination angles, they would appear redder in the NIR colors. Other factors may also be contributors to the observed diversity of ultrcool dwarfs' NIR photometry. Metallicity, one of the fundamental parameters that contribute to the physical properties of stars, is expected to also affect the photometry of ultracool dwarfs (\citealt{Burrows2001}, \citealt{Saumon2008}, \citealt{Fortney2008}, \citealt{Marley2021}). In this work, we aim to collect a sample of ultracool dwarfs with metallicity information to investigate the relationship between the metallicity and photometric properties of these objects.

Late-M (M6-M9) and L dwarfs are known to have cloudy atmospheres with silicate condensates that are optically thick in the NIR (e.g., \citealt{Burrows1999}, \citealt{Marley2001}, \citealt{Lodders2002}, \citealt{Allard2011}). These silicates then dissipate and settle deeper into the atmosphere as brown dwarfs cool into early-T dwarfs and later spectral types (e.g., \citealt{Marley2002}, \citealt{Lodders2006}, \citealt{Zhang2021}). Other species of clouds such as water, chlorides, and sulfides start to form in the cooler atmospheres of T dwarfs and later spectral types, but they are physically and optically thin in the NIR, making them effectively "cloudless" (\citealt{Visscher2006}, \citealt{Morley2012}, \citealt{Morley2014b}, \citealt{Lacy2023}). Therefore, mid/late-T dwarfs' NIR photometry is not affected by clouds the same way as the late-M and L dwarfs. Detailed studies of condensate formation and other related processes in ultracool dwarfs' atmospheres are still in progress (e.g., \citealt{Allard2014}, \citealt{Marley2015}). One possible cause for the variation in late-M and L dwarfs' NIR colors is that the amount of cloud formation is correlated with metallicity of the ultracool dwarfs.  

By establishing a sample of late-M and L dwarfs with known metallicities, we can investigate the imprints of atmospheric composition on the photometry of these objects. However, there is a paucity of detailed studies on metallicity properties for late-M and L dwarfs. The reason for this is measuring the metallicity of these cool objects is difficult due to their molecule-rich atmospheres. The many molecular lines/bands in their spectra hinder accurate extraction of ultracool dwarfs' metallicity information. Many strong efforts have been put into accurately measuring the metallicity of these objects by fitting low-temperature atmospheric models, such as \citet{Line2015}, \citet{Zalesky2019}, \citet{Zhang2021c}, \citet{Xuan2022}, and \citet{Hood2023}, etc. These measurements are subject to modeling systematics. Another common way to study the metallicity properties of the ultracool dwarfs is when they are companions of brighter stars (e.g. \citealt{Mann2014}). By measuring the metallicity of the host stars and assuming the primary and the secondary objects formed from the same molecular cloud, we can infer the metallicities of the ultracool companions from the metallicities of their host stars. This method provides us higher accuracy metallicity measurements of ultracool dwarf companions. We refer to these companions as `benchmark objects' given that some of their physical properties are known from independent information. 

In this work we focus on studying how metallicity is correlated with late-M and L dwarfs' observed near-IR colors, seeking to insight into whether metallicity affects the condensate formation in their atmospheres. We gathered a sample of late-M and L dwarf benchmarks (Section \ref{sec:data}), performed a correlation analysis for their $(J-K)$ color and metallicity (Section \ref{sec:analysis}), and compared these results with multi-metallicity atmospheric models (Section \ref{sec:dataModelComparison}). In addition, we discuss the possibility of using the empirical relationships we find as an easy and efficient method to provide bulk metallicity estimates for late-M and L dwarfs given their NIR photometry (Section \ref{sec:empiricalRelation}). 

\section{Sample Selection} \label{sec:data}
To study the correlation between metallicity and near-IR photometry for late-M and L dwarfs, we constructed a sample of benchmark late-M and L companions that have both $J-K$ colors and metallicity measurements of their host stars. $J-K$ was chosen because most M and L dwarfs have J and K photometry, making it a color commonly used in the study of ultracool dwarfs. As described below, we gathered late-M and L dwarf companions from the \textit{UltracoolSheet} \citep{UCS2020} and \citet{Mann2014}. Additionally, we identified 6 new benchmark ultracool dwarfs using the \citet{EB2021} binary catalog. Our total sample consisted of 67 objects, including 35 late-M dwarfs and 32 L dwarfs. 

\subsection{Previously Known Benchmarks} \label{subsec:UCSobj}
The $UltracoolSheet$ (UCS) is a library of 3000+ ultracool dwarfs with spectral types, photometry, astrometry, multiplicity, and companion information. The UCS provides both optical and IR spectral types when available. Following the field convention, the optical spectral types were used whenever available. Otherwise, the IR spectral types were used. For most objects in our selected sample with both optical and IR spectral types, their optical and IR spectral types are the same. There are only two objects with the largest discrepancy of half a spectral type between optical and IR. As for metallicity,  we obtained the benchmark objects' metallicities by finding their primary stars in APOGEE \citep{Apogee} and Hypatia \citep{Hypatia}. ASPCAP \citep{Garcia2016}, the spectral fitting pipeline for measuring chemical abundances in APOGEE, uses solar normalization from \citet{Asplund2005} for its metallicity measurements. To ensure consistency, we queried Hypatia metallicities with the same solar normalization. For objects that were not in APOGEE nor Hypatia, we checked SIMBAD for metallicity information from literature. In total, we gathered 17 late-M dwarf companions and 30 L dwarf companions with known metallicities from the UCS. Young objects and subdwarfs were excluded from the sample because they have different photometric properties compared to the field objects. 

\citet{Mann2014} contains a list of mid- to late-M dwarfs with known metallicities that they used to study metallicity calibrations for ultracool dwarfs. The metallicity measurements were carefully fitted and calibrated by using high resolution spectra and  Spectroscopy Made Easy (SME; \citealt{SME1996}). We included the late-M dwarfs in this study as part of our sample, which added an additional 11 late-M dwarfs with known metallicities to our total sample. \citet{Mann2014} provided the IR spectral types of these objects.

\subsection{Mining the El-Badry et al. (2021) binary catalog} \label{subsec:EBobj}

To increase our benchmark sample, we mined the \citet{EB2021} binary catalog for late-M dwarfs and L dwarfs. The \citet{EB2021} catalog contains 1.3 million spatially resolved binaries within $\sim$ 1 kpc of the Sun identified from Gaia eDR3 (\citealt{Gaia_eDR3}). We used \textit{Phototype}, a spectral energy distribution (SED) fitting method introduced by \citet{Skrzypek2015}, to identify late-M dwarfs and L dwarfs using photometry from large sky surveys. \emph{Phototype} summarizes a library of known objects as polynomial fits to colors as a function of spectral type (SpT). In our case, the UCS and \citet{Best2018catalog} was used to build the color functions with \textit{grizy} band data from Pan-STARRS 1 (PS1; \citealt{PS1_2016}), \textit{JHK} band data from the Mauna Kea observatories (MKO; \citealt{UHS_J2018}, \citealt{UHS_K2018}, \citealt{UKIDSS2007}), and \textit{W1} and \textit{W2} from CatWISE \citep{catWISE2020}. We used the $W1$ band as the reference filter to calculate the colors (e.g., $i - W1$, $W1 - W2$) of each UCS object (excluding binaries, subdwarfs and young objects), because it is available for 85\% of objects in our candidate list. The colors were fitted as polynomial functions for spectral types from M0 through Y0 (which corresponds to integers 0 through 30) and were weighted by the inverse square of the photometric errors (Table \ref{tab:colorModels}). For each color, only 5th order polynomials were fitted initially, we then successively increased the order as long as a polynomial with n+1 orders would bring a significantly improved fit (i.e., $\Delta \chi^2 > 7$) than the nth order polynomial. Then model templates were generated from the polynomial fits for each spectral type and least-squares fitting was performed on each secondary star in the \citet{EB2021} catalog with suitable photometry. A secondary star is then assigned the spectral type of the model template that results in the lowest $\chi^2$.

\begin{figure}
 \centering
 \includegraphics[width=0.83\textwidth]{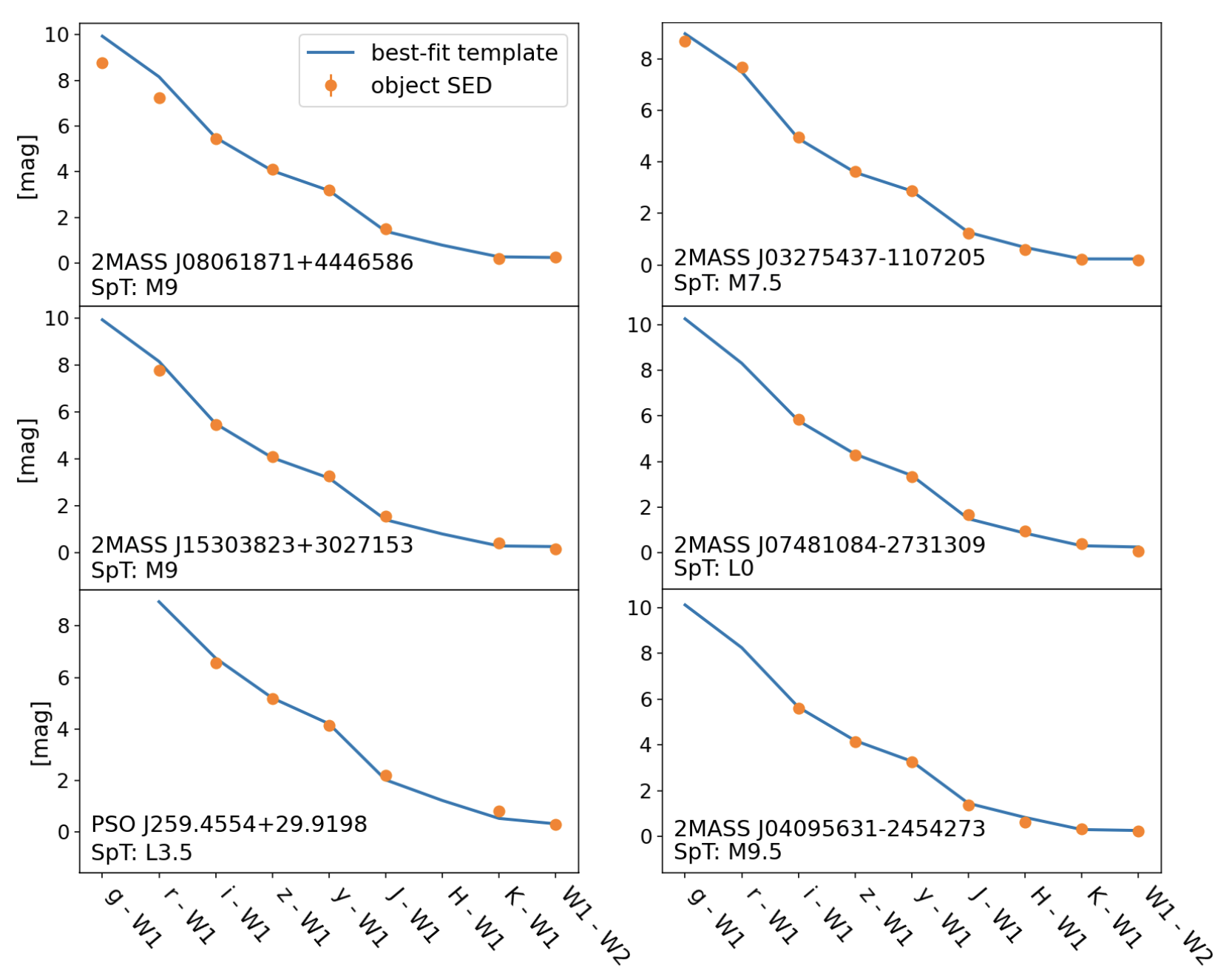}
 \caption{Spectral energy density of the 6 ultracool dwarfs identified from the \citet{EB2021} binary catalog that are not in the UCS. Error bars for the photometry are smaller than the symbol size.}
 \label{fig:ebObj}
\end{figure}

\begin{figure}
 \centering
 \includegraphics[width=0.5\textwidth]{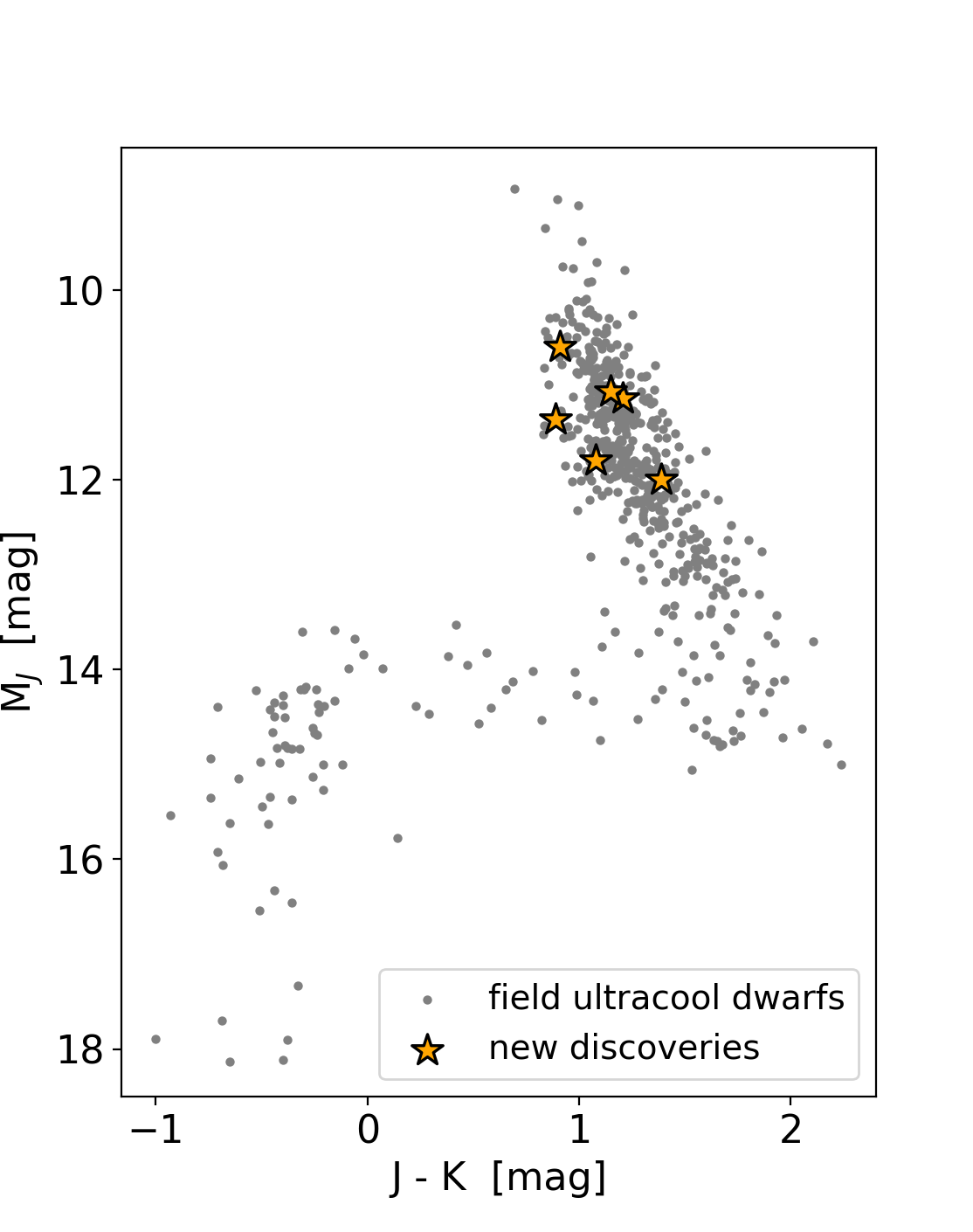}
 \caption{Color magnitude diagram of our 6 discovered ultracool companions (yellow stars) compared to field dwarfs in the UCS (grey dots).}
 \label{fig:ebCMD}
\end{figure}

Before spectral typing the secondary objects in the \citet{EB2021} binary catalog with the SED fitting algorithm, we first placed a distance limit of 200 pc on the catalog. The 200 pc limit was chosen because most L dwarfs should have at least 6 bands of photometry out to this distance, that could enable a robust spectral typing for \textit{Phototype}. Then we retrieved photometry of the companions from PS1 \citep{PS1_2016}, the Two Micron All Sky Survey (2MASS; \citealt{2MASS2006}), the UKIRT Hemisphere Survey (UHS; \citealt{UHS_J2018}), \textit{the UKIRT Infrared Deep Sky Survey} (UKIDSS; \citealt{UKIDSS2007}), and CatWISE \citep{catWISE2020}. For objects that have $JHK$ band data in the MKO photometric system, including UHS and UKIDSS, their MKO measurements are used instead of 2MASS. For objects with only 2MASS data, we convert their photometry to the MKO system using formulas from \citet{Stephens2004}, because our SED templates use the MKO system. We further limit the sample to objects that have at least 4 bands of photometry. 

Using the best-fit SpT from \textit{Phototype}, we compared our candidates' SpTs and absolute G-band magnitudes from Gaia \citep{Gaia_eDR3} to the G-band absolute magnitude versus SpT trend calculated from ultracool dwarfs in the UCS. Candidates with fitted SpT ranging from M9 to L9 that fell within $\pm 2$ mag of the known G-band trend were studied by looking at their PS1 image cutouts to confirm that the objects are point sources. Then, we examined the $\chi^2$ as a function of template SpT resulting from the SED fits, selecting objects with well-defined $\chi^2$ minima. We identified 14 ultracool dwarfs among these secondary stars, of which 6 are new discoveries (4 late-M dwarfs and 2 L dwarfs). Their SEDs correspond well to the best-fit SED templates (Figure \ref{fig:ebObj}). The 6 discoveries also land on the correct place on the color magnitude diagram (Figure \ref{fig:ebCMD}), further indicating that they have accurate spectral type estimates. Then, similar to the objects we gathered from the UCS, we crossmatched the host stars of these 6 objects to the APOGEE \citep{Apogee} and Hypatia \citep{Hypatia} catalogs to obtain their metallicity information.

Adding these 6 new objects to the sample gathered from the UCS and \citet{Mann2014}, we ended up with a final sample of 64 objects: 32 late-M dwarfs (Table \ref{tab:Mtab}) and 32 L dwarfs (Table \ref{tab:Ltab}). 

\section{Correlation Analysis} \label{sec:analysis}
\begin{figure}
 \centering
 \includegraphics[width=0.8\textwidth]{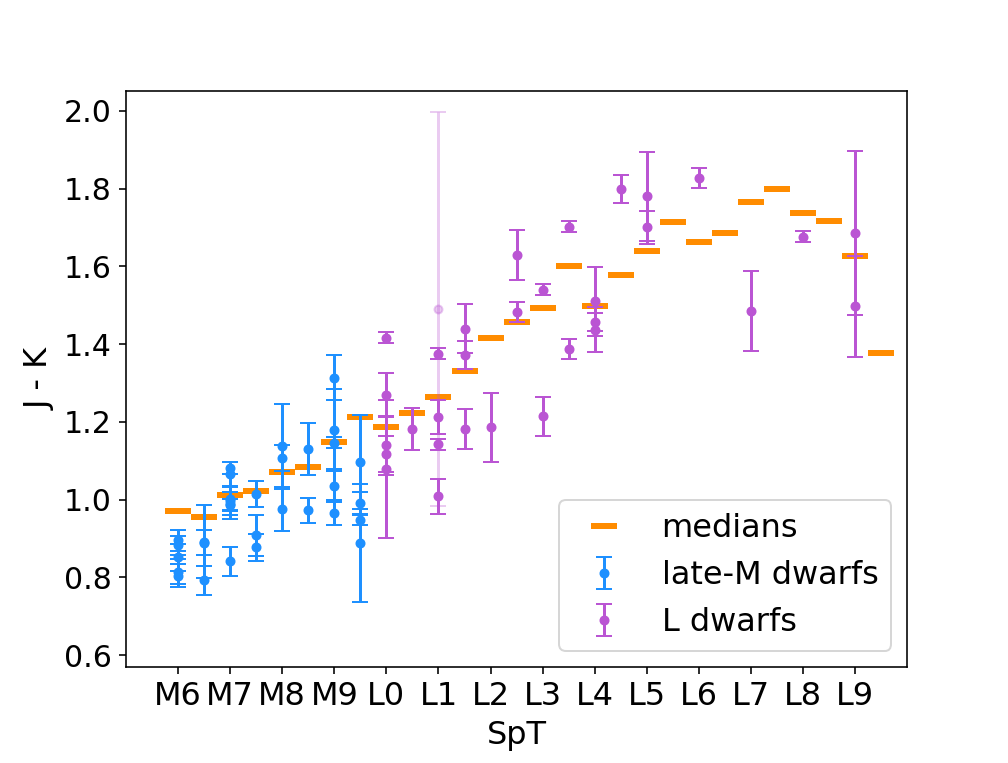}
 \caption{$J-K$ color of our late-M and L sample compared with the median $J-K$ colors of each SpT. Objects with signal to noise ratio (SNR) $< 3$ are plotted in a lighter shade.}
 \label{fig:JKvsSpT}
\end{figure}

\begin{figure}
 \centering
 \includegraphics[width=0.8\textwidth]{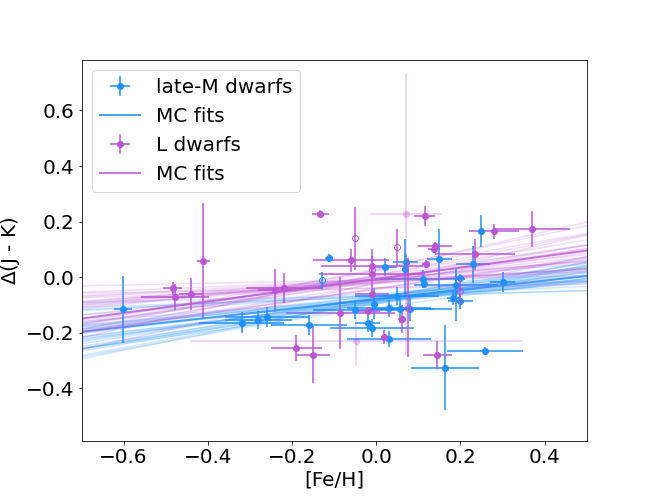}
 \caption{$\Delta (J-K)$ of all objects in the late-M/L sample as a function of [Fe/H]. Similar to Figure \ref{fig:JKvsSpT}, objects with SNR $< 3$ are plotted in lighter shades. Objects plotted with hollow markers do not have metallicity errors. The lines show 30 samples of the MC fits.}
 \label{fig:JKvsMet}
\end{figure}

As shown in Figure \ref{fig:ebCMD}, the $J-K$ colors of ultracool dwarfs follow a defined trend. However, different physical processes that take place in the atmospheres of these objects can cause color variations, making some objects appear redder or bluer than others. In our case, we want to understand how metallicity affects the photometry of ultracool dwarfs. Therefore, we calculated the $J-K$ color for all our benchmarks and compared them to the median $J-K$ color (Table \ref{tab:JKmed}) for each SpT (calculated using all objects in the UCS excluding binaries, subdwarfs, and young objects) using the MKO photometric system (Figure \ref{fig:JKvsSpT}). The difference between the $J-K$ color of each object and the median of its SpT, known and $J-K$ anomaly and denoted as $\Delta (J-K)$, was calculated and plotted as a function of [Fe/H] (Figure \ref{fig:JKvsMet}). 

Within our sample, there are 7 objects (2 late-M dwarfs and 5 L dwarfs) that have published metallicities without reported uncertainties. These objects were omitted in the correlation analysis, but they are still presented in Figure \ref{fig:JKvsMet} to show that there are no significant outliers in our sample. Hence, our correlation analysis was performed using the remaining 57 objects that have both $J-K$ color and metallicity. In the sample, 11 objects use metallicities from APOGEE, 17 from Hypatia, and 14 from \citet{Mann2014}. The remaining 15 objects use metallicity measurements from other various sources documented in Table \ref{tab:Mtab} and \ref{tab:Ltab}. 

As mentioned in Section \ref{sec:data}, APOGEE and Hypatia both use solar normalization from \citet{Asplund2005}. \citet{Mann2014} and APOGEE both use spectral fitting for measuring the stellar abundances. Metallicity measurements from the literature are not necessarily calibrated with each other, e.g. different sources most likely do not use the same normalization and methods, whether it be spectral fitting or curve of growth. The inconsistencies may inject extra systematic uncertainties in our analysis. \citet{Hypatia} found a median error of 0.04 dex across metallicities calibrated with different solar normalizations, which is a small effect compared to the metallicity range of our sample. Nonetheless, we made an effort to use metallicity measurements from APOGEE and Hypatia whenever possible. Due to these unaccounted uncertainties, the errors in our analysis represent the error floor. We performed the same correlation analysis for three sets of data: the total sample of 57 ultracool dwarfs, the 30 late-M dwarfs, and the 27 L dwarfs.

We used Spearman's rank correlation coefficient to examine whether the color anomaly $\Delta (J-K)$ and [Fe/H] are correlated. To account for measurement uncertainties, we simulated random samples of the data assuming Gaussian errors and ran 10,000 iterations of the test. A small fraction of the sample have implausibly small photometric errors ($\leq$ 0.01 mag) reported from UKIDSS, which were computed by direct error propagation of the flux error using the formula $\sigma_{mag} = \frac{2.5}{ln(10)}\frac{\sigma_f}{f}$. We set a floor to these photometric errors of 0.01 mag, which was motivated by fundamental calibration of UKIDSS to 2MASS, so objects with extremely small uncertainties do not skew our analysis. The total sample returned a Spearman's rank correlation coefficient (and standard deviation) of $0.26\pm0.07$ and its corresponding Student's t value resulted in a $97.5\% ^{+1.5\%}_{-7.5\%}$ confidence to reject the null hypothesis, namely that the parameters are not correlated with each other. The late-M dwarfs sample had a correlation coefficient of $0.31\pm0.10$ corresponding to a $95\%^{+2\%}_{-10\%}$ confidence level. And the L dwarfs sample resulted in a Spearman's rank correlation coefficient of $0.32\pm0.11$ and a $90\%^{+8\%}_{-5\%}$ confidence level. This suggests that both late-M and L dwarfs show correlation between $\Delta (J-K)$ and [Fe/H] with high confidence level. 

\begin{figure}
 \centering
 \includegraphics[width=0.7\linewidth]{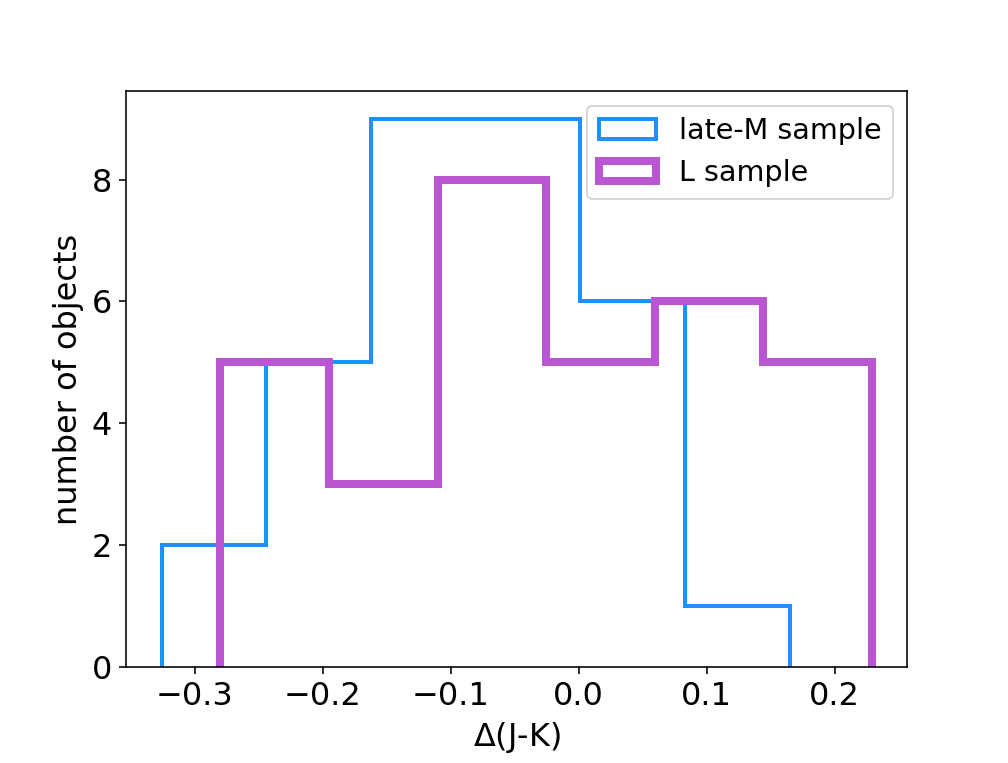}
 \caption{A histogram of our late-M dwarfs and L dwarfs as a function of color anomaly $\Delta(J-K)$.}
 \label{fig:JKhist}
\end{figure}

\begin{figure}
    \centering
    \includegraphics[width=0.7\linewidth]{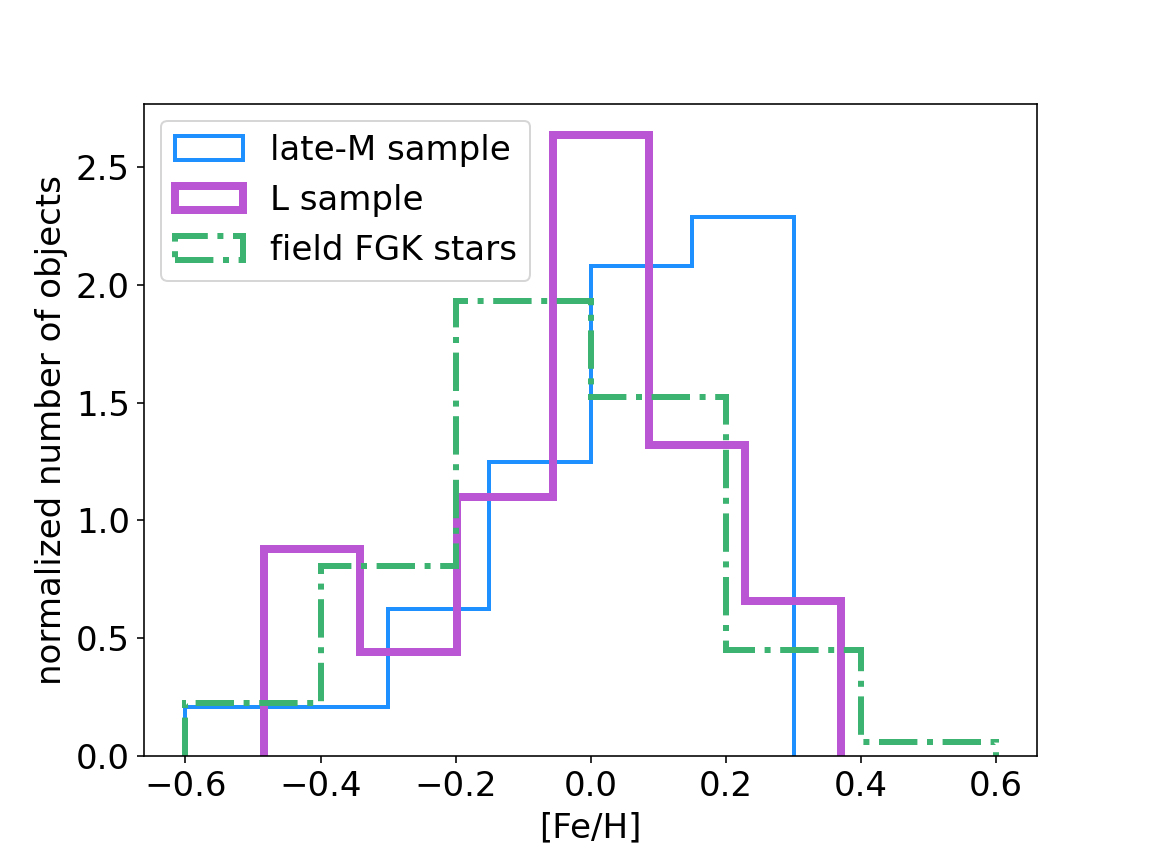}
    \caption{A normalized histogram of our late-M and L dwarfs as a function of metallicity compared to field FGK stars.}
    \label{fig:metHist}
\end{figure}

In attempt to summarize our observation and formulate an empirical relationship between $J-K$ anomaly and metallicity for late-M and L dwarfs, we applied 2 linear regression methods to fit the data. First, we drew random samples of the data assuming Gaussian errors for both $\Delta(J-K)$ and [Fe/H], then we used Ordinary Least Squares (OLS) to fit the drawn samples to linear models. This Monte Carlo (MC) process was repeated 1,000,000 iterations for the late-M dwarfs and the L dwarfs separately. The late-M dwarfs were best-fit to a slope of $0.17\pm0.07$ and a constant of $-0.08\pm0.01$; the L dwarfs resulted in a fitted slope of $0.20^{+0.07}_{-0.08}$ with a constant of $-0.01\pm0.02$ for the L-dwarfs $(J - K)$ anomaly and metallicity (Figure \ref{fig:JKvsMet}). This means for both populations, the more metal rich the objects appear to have redder $J-K$ colors. We speculate that the presence of more metals induces more condensate formation in the atmospheres of late-M and L dwarfs, thus making them appear redder in the NIR compared to objects of similar SpT but lower metallicity. In the following section, we show a comparison of our empirical findings with synthetic L dwarf spectra generated from Drift-Phoenix (\citealt{Dehn2007}; \citealt{Helling2008}; \citealt{Witte2009}) and discuss its significance.

We also experimented with using Orthogonal Distance Regression (ODR; \citealt{Boggs1989}) for fitting our data to linear models. Using ODR, the late-M dwarfs resulted in a slope of $0.52\pm0.16$ with a constant of $-0.07\pm0.03$; the L dwarfs were best fit to a slope of $0.48\pm0.19$ and a constant of $0.02\pm0.04$. Similar to the MC results, $\Delta(J-K)$ has a positive relationship with [Fe/H] for both late-M and L dwarfs, but the ODR fits have steeper slopes. 

One other detail to note here is that our late-M sample of benchmarks seems to have bluer $J-K$ anomaly than our L sample based on our MC fits. To better understand the populations that we are working with and test if the late-M sample is inherently bluer than the L sample, we used the 2-sample Kolmogorov-Smirnov (KS) test to study them. The 2-sample KS test is a standard statistical procedure to test whether two underlying one-dimensional probability distributions differ. The 2-sample KS test showed that we cannot confidently differentiate the late-M and L samples given their $\Delta(J-K)$ distribution (Figure \ref{fig:JKhist}) even though they seem bluer in Figure \ref{fig:JKvsMet}. We also performed the 2-sample KS test between our samples with field FGK stars from \citet{Casa2011} to assure that their metallicity distributions are consistent with the field stars (Figure \ref{fig:metHist}). The KS tests showed that both the late-M and L samples have metallicity distributions consistent with the field FGK stars.

\section{Comparison with model atmospheres}\label{sec:dataModelComparison}

\begin{figure}
 \centering
 \includegraphics[width=0.45\textwidth]{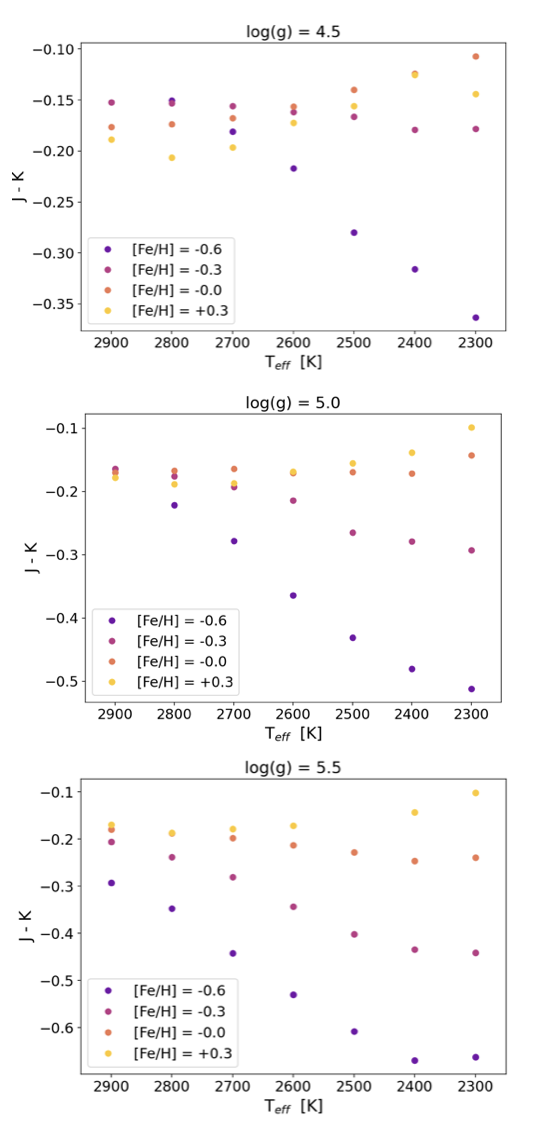}
 \caption{Drift-Phoenix models with a temperature range of [2300 - 2900 K] and a log(g) range of [4.5, 5.0, 5.5] log($\frac{cm}{s^2}$), which corresponds to late-M dwarfs. All photometry is on the MKO system.}
 \label{fig:dp_mod_M}
\end{figure}

\begin{figure}
 \centering
 \includegraphics[width=0.7\textwidth]{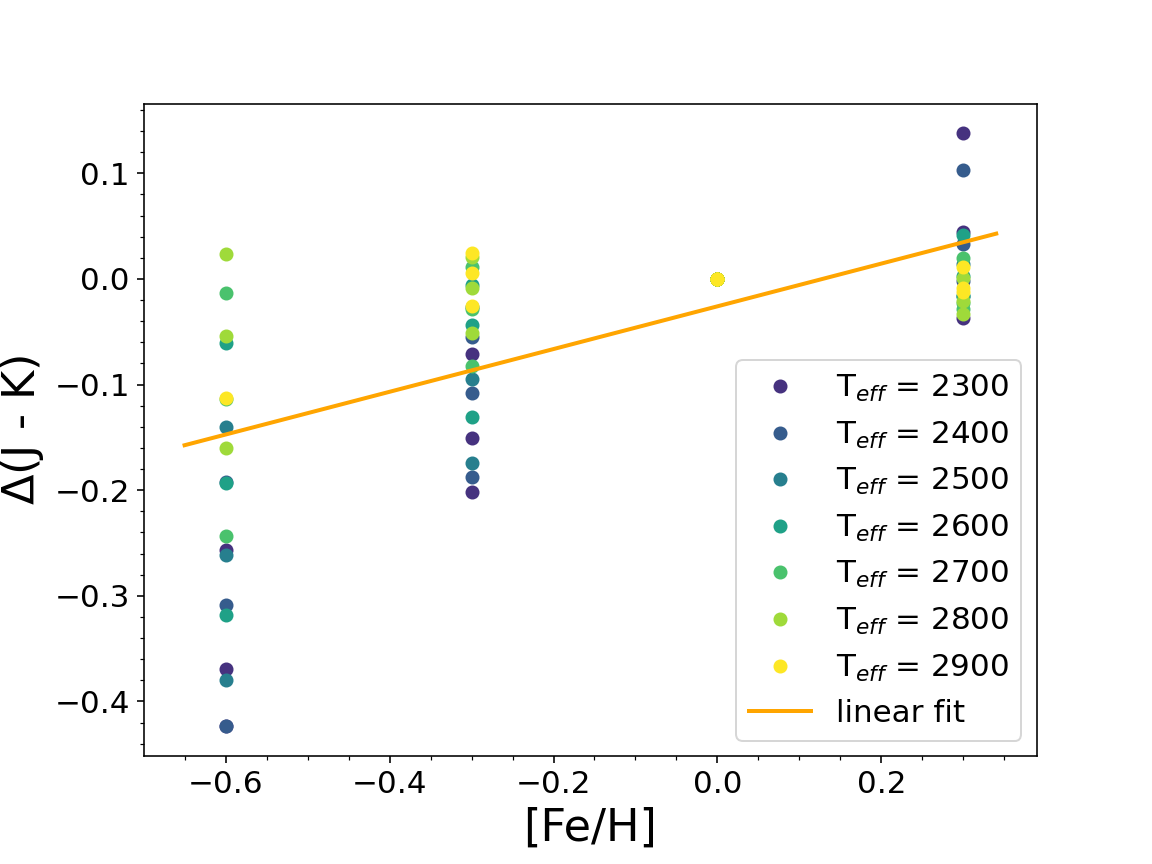}
 \caption{$\Delta$(J-K) of Drift-Phoenix models calculated with respect to the solar metallicity models as a function of [Fe/H], which serves as a comparison to our observational data in Figure \ref{fig:JKvsMet}. The temperature range is chosen to correspond to field late-M dwarfs. Note that because $\Delta(J-K)$ is calculated with respect to solar metallicity models, so models across all temperatures have $\Delta(J-K)$ of 0 for the solar metallicity models and the points are stacked on top of each other.}
 \label{fig:JKvsMet_mod_M}
\end{figure}

\begin{figure}
 \centering
 \includegraphics[width=0.45\textwidth]{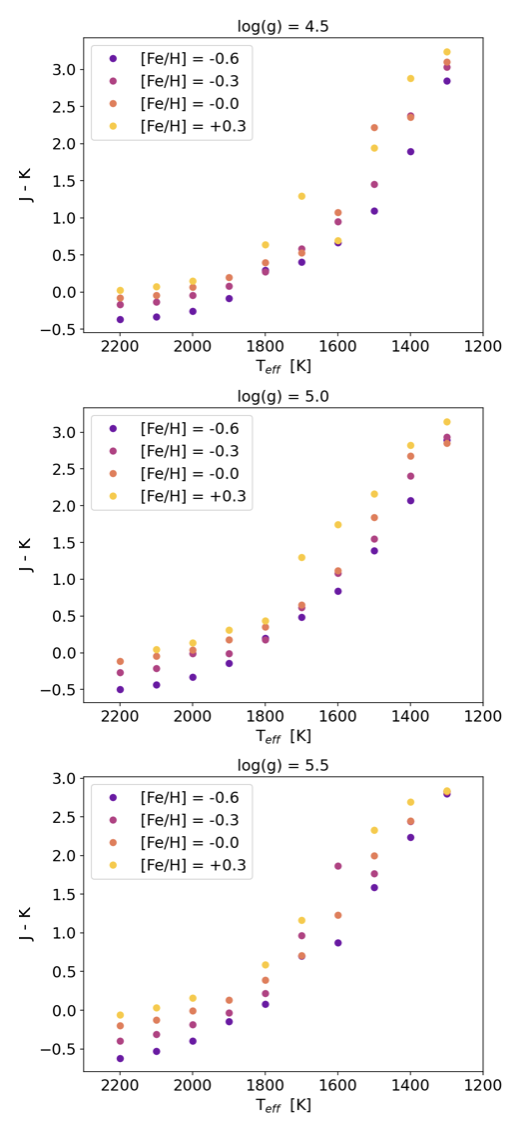}
 \caption{Drift-Phoenix models with a temperature range of [1300 - 2200 K] and a log(g) range of [4.5, 5.0, 5.5] log($\frac{cm}{s^2}$), which corresponds to L dwarfs. All photometry is on the MKO system.}
 \label{fig:dp_mod}
\end{figure}

\begin{figure}
 \centering
 \includegraphics[width=0.7\textwidth]{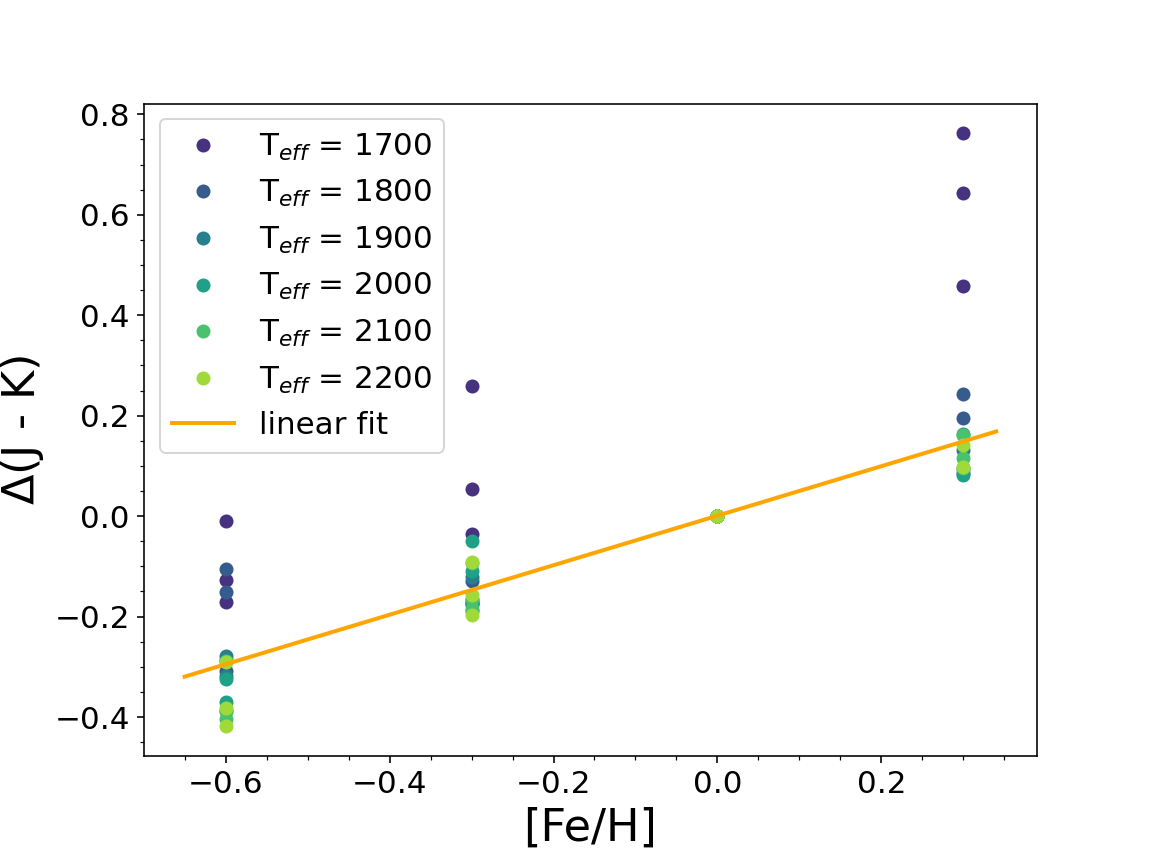}
 \caption{$\Delta$(J-K) of Drift-Phoenix models calculated with respect to the solar metallicity models as a function of [Fe/H], which serves as a comparison to our observational data in Figure \ref{fig:JKvsMet}. The temperature range is chosen to correspond to field L dwarfs. Similar to Figure \ref{fig:JKvsMet_mod_M}, because $\Delta(J-K)$ is calculated with respect to solar metallicity models, so models across all temperatures have $\Delta(J-K)$ of 0 for the solar metallicity models and the points are stacked on top of each other.}
 \label{fig:JKvsMet_mod}
\end{figure}

It is always valuable to compare observational data to simulated models because observations verify the assumptions that are used for building models. In the iterative process of taking observations, then making assumptions and building models that match the observations, we build more accurate models that better represent the real world. This is common practice to study phenomena that we cannot take direct measurements of. Here, we compare our observed results to the Drift-Phoenix model atmospheres. Drift-Phoenix is a model atmosphere code derived from the PHOENIX code family (\citealt{Hauschild1999}, \citealt{Baron2003}) and the DRIFT models (\citealt{Woitke2003, Woitke2004}; \citealt{Helling2006}; \citealt{Helling2008b}) that simulate stellar/planetary atmospheres including the formation of condensates. Additionally, this set of models is multi-metallicity while most other models that take cloud formation into consideration are not. Hence the Drift-Phoenix models are relevant to this work. Note that the model atmospheres do not fit actual spectra of these objects well, we are only using them in a differential case for color calculations. To study how our observed trend compares with theoretical atmospheric models, we calculated the $J-K$ color of the models: with $T_{eff}$ ranging from 2300 K to 2900 K for the late-M dwarfs (Figure \ref{fig:dp_mod_M}) and $T_{eff}$ ranging from 1300 K to 2200 K for the L dwarfs (Figure \ref{fig:dp_mod}); log(g) of 4.5, 5.0, and 5.5 log($\frac{cm}{s^2}$); and metallicities ranged from -0.6 to 0.3 with increments of 0.3. The higher temperature range covers late-M dwarfs while the lower temperature range covers L dwarfs, and the gravity range covers field objects while leaving out young objects. Considering that the late-L dwarfs entering the transient phase of the L-T transition that is not accurately modeled, we limit the temperature range to $\geq 1700$ K for the following analysis, which corresponds to SpT earlier than L6 based on the $T_{eff}$-SpT relationship in \citet{Dupuy2017}. 

Given the observational data shows a value close to zero for $\Delta(J-K)$ at solar metallicity, we subtracted the modeled $J-K$ colors by the solar metallicity models and plotted the results as a function of metallicity in Figure \ref{fig:JKvsMet_mod_M} and Figure \ref{fig:JKvsMet_mod} to compare to Figure \ref{fig:JKvsMet}. For the late-M dwarfs, the models resulted in a slope of $0.20\pm0.03$ and a constant of $-0.03\pm0.01$ using OLS linear regression. We can see in Figure \ref{fig:JKvsMet_mod_M} that the fit is heavily influenced by the widespread data in $\Delta(J-K)$ of various $T_{eff}$ and log(g). This wide variation is pulling the y-intercept below zero, which is somewhat evocative of the late-M dwarfs' observational data which has a fitted y-intercept of $-0.08\pm0.01$ (MC approach). Looking at the fitted slopes, the models' result overlaps with our observational result of $0.17\pm0.07$, suggesting that the current Drift-Phoenix models are reasonable representations of late-M dwarfs looking from the perspective of this metallicity distribution. For the L dwarfs, the models also show a positive linear relationship (fitted using OLS) with a slope of $0.49\pm0.02$ and a constant of $0.001\pm0.008$. The slope is a much larger value compared to our MC observational value of $0.20^{+0.07}_{-0.08}$ but in the same direction. 

\section{Empirical relationships for photometric metallicities of late-M and L dwarfs} \label{sec:empiricalRelation}
Past studies have found empirical relationships between $J-K$ color and metallicity of M-dwarfs, which can provide a convenient way for obtaining the metallicity of objects from photometry alone. For example, partly based on the color-metallicity relationship noted in \citet{Leggett1992} and \citet{Lepine2005}, \citet{Johnson2012} found an empirical relationship between metallicity and $J-K$ using a volume-limited M dwarf sample:
\begin{equation}
    [Fe/H] = -0.050 + 3.520\Delta(J-K)
\end{equation}
where $\Delta(J-K)$ is defined as
\begin{equation*}
    \Delta(J-K) = \left\{
        \begin{array}{ll}
            (J-K) - 0.835 & \quad (V-K) < 5.5 \\
            (J-K) - \Sigma_{i=0}a_i(V-K)^i& \quad (V-K) \geq 5.5
        \end{array}
    \right.
\end{equation*}
with $a  = [1.637, -0.2910, 0.022557$]. This relationship is valid for stars with $-0.1<\Delta(J-K)<0.1$ and $V-K > 3.8$, and yielded metallicities accurate to $\pm0.15$ dex based on their sample. 

\citet{Mann2013} tested this relationship with a different sample of early- to mid-M dwarfs and found an RMS of 0.20 dex about the fitted relation. In attempt to improve the calibration, their best-fit relation was
\begin{equation}
    [Fe/H] = -0.11 + 3.14\Delta(J-K)
\end{equation}
which slightly improved the RMS to 0.19 dex. 

In this work, by rearranging the best-fit relationships in Section \ref{sec:analysis}, we obtain two separate metallicity relations as functions of $\Delta(J-K)$ for late-M dwarfs and L dwarfs. For late-M dwarfs, the relation is
\begin{equation}
    [Fe/H]_M = 0.48 + 5.99\Delta(J-K)_M
\end{equation}
and for L dwarfs, the relation is
\begin{equation}
    [Fe/H]_L = 0.03 + 4.90\Delta(J-K)_L.
\end{equation}
Of course, we define our $\Delta(J-K)$ differently than \citet{Johnson2012}, being as the difference between an object's $J-K$ color and the median $J-K$ color of its corresponding SpT (Table \ref{tab:JKmed}). The residual error to our fit for the late-M dwarf sample has an RMS of 0.61 dex, and 0.95 dex for L dwarf sample. The large residual errors may be attributed to the fact that metallicity is not the only thing affecting $J-K$ colors of these objects (e.g. viewing geometry and surface gravity as discussed in Section \ref{sec:intro}), which leads to large scatter in the fitted relations. Inconsistencies of methods that were used for metallicity measurements in different literature that we collected our sample from also serve as a contributor to the errors. Thus, these empirical relationships are not reliable for metallicity measurements for late-M and L dwarfs but may serve as an easy way to obtain crude metallicity information when only the photometry is available. 

\section{Conclusion} \label{sec:conclusion}
We assembled a sample of 64 ultracool dwarf companions with known metallicities from their primary stars, comprising 32 late-M dwarfs and 32 L dwarfs. This included 6 new ultracool dwarfs that we identified using the binary catalog from \citet{EB2021} and SED fitting. Spearman's rank correlation test showed that the late-M dwarfs sample (the L dwarfs sample) has a positive correlation with 95\% (90\%) confidence level between the color anomaly $\Delta(J-K)$ and metallicity. A linear fit to the data gives a slope of $0.17\pm0.07$ ($0.20^{+0.07}_{-0.08}$) and a constant of $-0.08\pm0.01$ ($-0.01\pm0.02$). The more metal-rich an object is, the redder it appears, which suggests that higher metallicity induces more condensate formation in the atmospheres of these ultracool dwarfs. The Drift-Phoenix models also showed similar trends, but the L dwarf models had a larger slope of $0.49\pm0.02$. This suggests that the atmosphere models of these ultracool objects need to be improved to match observational data. We also used 2-sample KS test to confirm that our sample of ultracool dwarfs have consistent metallicity distribution compared to the field FGK stars.

This study serves to provide an observed relationship between $\Delta(J-K)$ color and metallicity for these objects to improve future atmospheric models, thus gain a better and more complete understanding of condensate formation in ultracool dwarf and giant exoplanet atmospheres. We also discuss the possibility to use the empirical relationship to obtain crudely approximated metallicity given a $(J-K)$ color. We may improve and further constrain the relationships found in this work using larger samples of late-M and L dwarfs, such as those that will result from the Legacy Survey of Space and Time (LSST) conducted at Rubin Observatory.

\facilities{Pan-STARRS, UKIRT}
%

\vspace{5mm}

\begin{longrotatetable}
\begin{deluxetable}{cccccccccccc}

\tabletypesize{\scriptsize}
\tablewidth{0pt}

\tablecaption{ Fitted polynomial coefficients of ultracool dwarf colors as functions of spectral types\label{tab:colorModels}}

\tablehead{
\colhead{Color} & \colhead{SpT} & \colhead{a$_0$} & \colhead{a$_1$} & \colhead{a$_2$} & \colhead{a$_3$} & \colhead{a$_4$} & \colhead{a$_5$} & \colhead{a$_6$} & \colhead{a$_7$} & \colhead{a$_8$} & \colhead{a$_9$}
}

\startdata
$g - W1$ & [0,11] & +4.412E+0 & +5.6163E-1 & -2.6113E-1 & +8.9422E-2 & -1.0354E-2 & +4.8532E-4 & -8.0770E-6 & & & \\
$r - W1$ & [0,16] & +3.228E+0 & +3.78073E-1 & -4.35955E-2 & -1.40900E-2 & +1.41348E-2 & -2.61549E-3 & +1.95128E-4 & -6.08834E-6 & +5.69910E-8 & \\
$i - W1$ & [0,20] & +2.553E+0 & +1.38372E-1 & +3.74034E-2 & -4.76123E-2 & +2.06378E-2 & -3.72891E-3 & +3.46061E-4 & -1.74675E-5 & +4.57227E-7 & -4.87456E-9\\
$z - W1$ & [0,28] & +2.236E+0 & +1.82857E-1 & -1.39936E-1 & +5.37396E-2 & -8.44757E-3 & +7.12631E-4 & -3.42884E-5 & +9.18231E-7 & -1.22666E-8 & +5.89381E-11\\
$y - W1$ & [0,29] & +2.063E+0 & +1.29408E-1 & -1.10970E-1 & +4.42354E-2 & -7.69674E-3 & +7.41032E-4 & -4.13195E-5 & +1.311955E-6 & -2.189699E-8 & +1.487118E-10\\
$J - W1$ & [0,30] & +9.366E-1 & +7.36787E-2 & -7.03136E-2 & +2.81689E-2 & -5.05757E-3 & +4.90706E-4 & -2.66342E-5 & +7.964306E-7 & -1.213588E-8 & +7.292408E-10\\
$H - W1$ & [0,30] & +2.736E-1 & +1.08221E-1 & -7.95243E-2 & +3.05663E-2 & -5.51001E-3 & +5.46211E-4 & -3.11239E-5 & +1.011902E-6 & -1.747484E-8 & +1.249830E-10\\
$K - W1$ & [0,30] & +9.896E-2 & +7.51401E-3 & +1.67941E-3 & +7.78834E-4 & -2.80064E-4 & +3.45669E-5 & -1.87863E-6 & +4.539632E-8 & -3.906790E-8 & \\
$W1 - W2$ & [0,30] & -2.732E-2 & -3.8613E-2 & -7.56159E-3 & +3.15515E-3 & -4.12880E-4 & +2.57523E-5 & -7.80082E-7 & +9.018347E-9 &  & \\
\enddata

\vspace{-0.5cm}

\end{deluxetable}
\end{longrotatetable}


\tabletypesize{\scriptsize}
\startlongtable
\begin{deluxetable}{cc|cc}
\tabletypesize{\scriptsize}

\tablecaption{\label{tab:JKmed}Median $J-K$ colors as a function of SpT calculated from field dwarfs in the UCS} 

\tablehead{
\colhead{\hspace{.75cm}SpT}\hspace{.75cm} & \colhead{\hspace{.75cm}$J-K$}\hspace{.75cm} & \colhead{\hspace{.75cm}SpT}\hspace{.75cm} & \colhead{\hspace{.75cm}$J-K$}\hspace{.75cm} 
} 
\startdata
    6.0 & 0.979 & 19.0 & 1.627\\
    6.5 & 0.956 & 19.5 & 1.376\\
    7.0 & 1.013 & 20.0 & 1303\\
    7.5 & 1.021 & 20.5 & 1.187\\
    8.0 & 1.072 & 21.0 & 1.071\\
    8.5 & 1.084 & 21.5 & 1.002\\
    9.0 & 1.149 & 22.0 & 0.827\\
    9.5 & 1.214 & 22.5 & 0.649\\
    10.0 & 1.188 & 23.0 & 0.426\\
    10.5 & 1.222 & 23.5 & 0.288\\
    11.0 & 1.263 & 24.0 & 0.135\\
    11.5 & 1.332 & 24.5 & -0.065\\
    12.0 & 1.416 & 25.0 & -0.248\\
    12.5 & 1.456 & 25.5 & -0.246\\
    13.0 & 1.493 & 26.0 & -0.387\\
    13.5 & 1.601 & 26.5 & -0.320\\
    14.0 & 1.497 & 27.0 & -0.282\\
    14.5 & 1.578 & 27.5 & -0.466\\
    15.0 & 1.640 & 28.0 & -0.430\\
    15.5 & 1.713 & 28.5 & -0.387\\
    16.0 & 1.664 & 29.0 & -0.556\\
    16.5 & 1.685 & 29.5 & -0.733\\
    17.0 & 1.764 & 30.0 & -0.910\\
    17.5 & 1.798 &&\\
    18.0 & 1.737 &&\\
    18.5 & 1.717 &&\\
\enddata 
\end{deluxetable}

\begin{longrotatetable}
\begin{deluxetable}{lllllllllllllllll}

\tabletypesize{\tiny}
\tablewidth{0pt}

\tablecaption{ M Dwarfs Sample\label{tab:Mtab}}

\tablehead{
\colhead{Name} & \colhead{RA} & \colhead{Dec} & \colhead{SpT}&  \colhead{J} & \colhead{K} & \colhead{$\Delta$(J-K)} & \colhead{Primary Star} & \colhead{P-SpT} & \colhead{Sep} & \colhead{Sep} & \colhead{[Fe/H]} & \colhead{References}\\
& \colhead{(J2000)} & \colhead{(J2000)} & & \colhead{[mag]} & \colhead{[mag]} & \colhead{[mag]} & & & \colhead{[as]} & \colhead{[AU]} & &
}

\startdata 
HIP 6217C & 19.9361 & 0.1051 & 9.5; IR & 15.402$\pm$0.006 & 14.454$\pm$0.008 & -0.266$\pm$0.010 & HIP 6217 & G6 & 27.4 & 3014.0 & 0.258$\pm$0.09 & 1, 1, 2, 2, 3 \\
2MASS J03184214+0828002 & 49.6756 & 8.4667 & 7.0; IR & 13.647$\pm$0.027 & 12.662$\pm$0.023 & -0.028$\pm$0.035 & NLTT 10534 & M3 &  75.5 & 3098.7 & 0.19$\pm$0.10 & 4, 4, 5, 5, 4 \\
2MASS J03275437-1107205 & 51.9767 & -11.1223 & [7.5] & 14.227$\pm$0.037 & 13.319$\pm$0.039 & -0.113$\pm$0.054 & UCAC3 158-9098 &  & 100.3 & 5327.0 & 0.056$\pm$0.08 & 6, 0, 5, 5, 3 \\
2MASS J04095631-2454273 & 62.4849 & -24.9076 & [9.5] & 15.807$\pm$0.085 & 14.919$\pm$0.126 & -0.326$\pm$0.152 & TYC 6456-1296-1 &  & 145.3 & 11706.0 & 0.163$\pm$0.08 & 6, 0, 5, 5, 3 \\
2MASS J04305157-0849007 & 67.7149 & -8.8169 & 8.5; IR & 12.739$\pm$0.022 & 11.767$\pm$0.023 & -0.113$\pm$0.032 & LP 655-23 & M4 &  14.4 & 450.0 & 0.030$\pm$0.10 & 7, 4, 5, 5, 4 \\
2MASS J07394386+1305070 & 114.9327 & 13.0852 & 8.0; opt & 13.910$\pm$0.096 & 12.773$\pm$0.052 & 0.065$\pm$0.109 & BD+13 1727 & K5 &  10.5 & 517.6 & 0.150$\pm$0.03 & 8, 9, 5, 5, 4 \\
2MASS J08061871+4446586 & 121.5780 & 44.7828 & [9.0] & 15.856$\pm$0.067 & 14.677$\pm$0.083 & 0.030$\pm$0.107 & 2M08061929+4446362 & &  23.2 & 1955.3 & 0.069$\pm$0.01 & 6, 0, 5, 5, 10 \\
NLTT19472 & 126.2143 & -3.6836 & 6.0; IR & 11.418$\pm$0.026 & 10.604$\pm$0.019 & -0.156$\pm$0.028 & HIP 41211 & F6 & 356.4 & 9729.7 & -0.28$\pm$0.08 & 11, 4, 5, 5, 12 \\
LSPM J0942+2351 & 145.7283 & 23.8553 & 6.5; IR & 13.146$\pm$0.024 & 12.257$\pm$0.021 & -0.067$\pm$0.032 & NLTT 22411 & M1 & 132.6 & 4759.6 & 0.05$\pm$0.10 & 13, 4, 5, 5, 4 \\
HIP 49046B & 150.1488 & 27.2849 & 6.5; IR & 12.930$\pm$0.092 & 12.038$\pm$0.021 & -0.064$\pm$0.094 & HIP 49046 & M0.5 & 136.1 & 4868.3 & 0.190 & 1, 1, 1, 5, 14 \\
20 LMi B & 150.2091 & 31.9289 & 6.0; IR & 10.125$\pm$0.018 & 9.243$\pm$0.018 & -0.088$\pm$0.025 & HIP 49081 & G3 & 158.2 & 2357.2 & 0.200$\pm$0.03 & 1, 4, 5, 5, 4 \\
PM I11055+4331 & 166.3788 & 43.5214 & 6.5; IR & 8.602$\pm$0.026 & 7.810$\pm$0.026 & -0.164$\pm$0.037 & HIP 54211 & M1 & 31.4 & 154.0 & -0.32$\pm$0.10 & 4, 4, 5, 5, 4 \\
NLTT28453 & 176.3975 & -20.3511 & 6.0; IR & 11.597$\pm$0.026 & 10.745$\pm$0.023 & -0.118$\pm$0.028 & HIP 57361 & M2.5 & 15.4 & 297.9 & -0.05$\pm$0.10 & 1, 4, 5, 5, 4 \\
2MASS J12003292+2048513 & 180.1371 & 20.8143 & 7.0; opt & 12.862$\pm$0.021 & 11.861$\pm$0.022 & -0.012$\pm$0.030 & G 121-42 & M3 &  204.0 & 5022.5 & -0.130 & 15, 15, 5, 5, 16 \\
HIP 59310B & 182.5412 & 18.9690 & 7.0; IR & 13.672$\pm$0.027 & 12.676$\pm$0.024 & -0.017$\pm$0.036 & HIP 59310 & K3  & 82.1 & 3735.5 & 0.30$\pm$0.03 & 1, 1, 17, 17, 18 \\
NLTT 30510B & 185.5767 & 36.7302 & 9.5; IR & 15.923$\pm$0.080 & 14.826$\pm$0.090 & -0.116$\pm$0.120 & NLTT 30510 & & 19.6 & 1352.4 & -0.601$\pm$0.02 & 1, 1, 17, 17, 10 \\
2MASS J13204159+0957506 & 200.1733 & 9.9639 & 7.0; IR & 13.650$\pm$0.010 & 12.584$\pm$0.029 & 0.053$\pm$0.031 & HIP 65133 & K2 &  168.0 & 6078.5 & 0.07$\pm$0.03 & 19, 4, 20, 5, 4 \\
HIP 65706B & 202.0872 & 30.0552 & 7.0; IR & 13.246$\pm$0.001 & 12.258$\pm$0.002 & -0.025$\pm$0.002 & HIP 65706 & K5  & 52.6 & 2227.6 & 0.114$\pm$0.01 & 1, 1, 2, 2, 11 \\
NLTT 36549 & 213.0505 & -0.5878 & 7.5; IR & 12.943$\pm$0.024 & 12.066$\pm$0.026 & -0.144$\pm$0.035 & NLTT 36548 & M3.5 & 12.0 & 321.6 & -0.26$\pm$0.10 & 21, 4, 5, 5, 4 \\
Gl 569B & 223.6226 & 16.1024 & 8.5; opt & 10.750$\pm$0.060 & 9.620$\pm$0.030 & 0.046$\pm$0.067 & Gl 569 A & M3 & 5.0 & 49.7 & 0.230$\pm$0.04 & 22, 23, 24, 24, 25 \\
2MASS J15303823+3027153 & 232.6586 & 30.4540 & [9.0] & 15.520$\pm$0.010 & 14.373$\pm$0.008 & -0.0026$\pm$0.013 & 2M15304088+3027059 & K5 & 35.5 & 2865.5 & 0.198$\pm$0.01 & 6, 0, 26, 27, 10 \\
2MASS J15575569+5914232 & 239.4821 & 59.2398 & 9.0; IR & 14.156$\pm$0.031 & 13.120$\pm$0.030 & -0.113$\pm$0.043 & HIP 78184 & K5 & 119.0 & 3871.0 & 0.080$\pm$0.10 & 7, 28, 5, 5, 4 \\
HIP 78916B & 241.6345 & 22.8927 & 8.0; IR & 15.148$\pm$0.018 & 14.172$\pm$0.053 & -0.096$\pm$0.056 & HIP 78916 & G0 & 35.5 & 3035.2 & -0.005$\pm$0.01 & 1, 1, 1, 17, 10 \\
HIP 81910B & 250.9563 & -26.8109 & 6.0; IR & 12.400$\pm$0.010 & 11.505$\pm$0.025 & -0.075$\pm$0.027 & HIP 81910 & G2.5 & 26.7 & 1254.1 & 0.183$\pm$0.02 & 1, 1, 1, 17, 29 \\
vB 8 & 253.8970 & -8.3944 & 7.0; opt & 9.632$\pm$0.029 & 8.791$\pm$0.023 & -0.172$\pm$0.037 & HIP 82817 & M3.5 & 231.0 & 1500.3 & -0.16$\pm$0.09 & 30, 30, 5, 5, 31 \\
LSPM J1717+5925B & 259.3792 & 59.4254 & 9.0; IR & 16.509$\pm$0.018 & 15.195$\pm$0.054 & 0.165$\pm$0.057 & LSPM J1717+5925 & G6 &  14.4 & 1762.6 & 0.250$\pm$0.00 & 1, 1, 1, 17, 10 \\
vB 10 & 289.2400 & 5.1504 & 7.5; IR & 9.760$\pm$0.025 & 8.745$\pm$0.022 & -0.0061$\pm$0.033 & HIP 94761 & M3 & 75.2 & 443.7 & 0.110$\pm$0.10 & 32, 4, 5, 5, 4 \\
2MASS J20103539+0634367 & 302.6475 & 6.5767 & 9.0; IR & 12.389$\pm$0.024 & 11.423$\pm$0.021 & -0.183$\pm$0.032 & NLTT 48838 & M4 & 142.8 & 2313.4 & -0.010$\pm$0.10 & 33, 4, 5, 5, 4 \\
LSPM J2049+3216W & 312.3073 & 32.2808 & 6.0; IR & 11.633$\pm$0.023 & 10.829$\pm$0.019 & -0.166$\pm$0.030 & HIP 102766 & K2.5  & 34.4 & 801.5 & -0.020$\pm$0.03 & 4, 4, 5, 5, 4 \\
LSPM J2153+1157B & 328.4458 & 11.9633 & 7.0; IR & 14.450$\pm$0.011 & 13.368$\pm$0.011 & 0.069$\pm$0.016 & LSPM J2153+1157 & & 11.3 & 601.2 & -0.112$\pm$0.01 & 1, 1, 1, 1, 10 \\
2MASS J22264440-7503425 & 336.6851 & -75.0618 & 8.0; opt & 12.353$\pm$0.023 & 11.246$\pm$0.023 & 0.035$\pm$0.033 & HD 212168 & G0 &  265.0 & 6222.2 & 0.020$\pm$0.03 & 34, 19, 5, 5, 35 \\
G 216-7B & 339.3856 & 39.3775 & 9.5; opt & 13.173$\pm$0.022 & 12.183$\pm$0.019 & -0.224$\pm$0.030 & HIP 111685 & M0 & 33.6 & 705.2 & 0.030$\pm$0.10 & 36, 36, 5, 5, 4 \\
\enddata

\tablecomments{2MASS J00034227-2822410 was considered for the sample, but it was not included due to its higher luminosity (by $>$ 2$\sigma$) compared to the field ultracool dwarfs of the same spectral type, suggesting that it might be a young object. The numbering for the spectral types follows the ultracool dwarfs convention such that M6 is labeled as 6.0, M7 is 7.0, so on and so forth. The type of SpT, whether is be optical or IR, is indicated with either the `opt' flag or the `IR' flag. Objects with spectral types with brackets around them are new benchmark objects we identified in this work. Spectral types are estimated from our SED fitting procedure. `P-SpT' stands for primary star SpT.\\
Reference format: first number is the discovery reference, second is the SpT reference, third and fourth are the J and K photometry reference, and the last number is the metallicity reference.
References: 0 = this work, 1 = \citealt{Deac14b}, 2 = \citet{Lawr12}, 3 = \citet{Steinmetz2020}, 4 = \citet{Mann2014}, 5 = \citet{Cutr03}, 6 = \citet{EB2021}, 7 = \citet{Cruz03}, 8 = \citet{Lepi02b}, 9 = \citet{Cruz07}, 10 = \citet{Apogee}, 11 = \citet{Alle2012}, 12 = \citet{Casa2011}, 13 = \citet{West11}, 14 = \citet{Gaid14}, 15 = \citet{Gizi00c}, 16 = \citet{Newt14}, 17 = \citet{Best2020b}, 18 = \citet{Grieves2018}, 19 = \citet{Reid08b}, 20 = \citet{Smit14a}, 21 = \citet{Bess91b}, 22 = \citet{Skru87b}, 23 = \citet{Henr90}, 24 = \citet{Dupu10}, 25 = \citet{Hypatia}, 26 = \citet{UHS_J2018}, 27 = \citet{UHS_K2018}, 28 = \citet{Thom13}, 29 = \citet{Liu2020}, 30 = \citet{Henr04}, 31 = \citet{Kuzn19}, 32 = \citet{Kirk91}, 33 = \citet{Luhm12d}, 34 = \citet{Luyt79b}, 35 = \citet{Ghezzi2010}, 36 = \citet{Schm07}. All parallax related information comes from \citet{GaiaDR2} and \citet{Gaia_eDR3}.
}

\end{deluxetable}
\end{longrotatetable}

\begin{longrotatetable}
\begin{deluxetable}{lllllllllllllllll}

\tabletypesize{\tiny}
\tablewidth{0pt}

\tablecaption{ L Dwarfs Sample\label{tab:Ltab}}
\tablehead{
\colhead{Name} & \colhead{RA} & \colhead{Dec} & \colhead{SpT}&  \colhead{J} & \colhead{K} & \colhead{$\Delta$(J-K)} & \colhead{Primary Star} & \colhead{P-SpT} & \colhead{Sep} & \colhead{Sep} & \colhead{[Fe/H]} & \colhead{References}\\
& \colhead{(J2000)} & \colhead{(J2000)} & & \colhead{[mag]} & \colhead{[mag]} & \colhead{[mag]} & & & \colhead{[as]} & \colhead{[AU]} & &
}

\startdata 
LHS 102B & 1.1453 & -40.7351 & 15.0; opt & 13.060$\pm$0.030 & 11.360$\pm$0.030 & 0.060$\pm$0.042 & GJ 1001A & M4 & 18.6 & 226.5 & -0.060$\pm$0.09 & 39, 37, 38, 38, 31 \\
NLTT 1011B & 4.8865 & 40.3159 & 12.0; IR & 15.485$\pm$0.059 & 14.299$\pm$0.066 & -0.230$\pm$0.089 & NLTT 1011 & & 58.5 & 3276.0 & -0.048$\pm$0.39 & 1, 1, 17, 17, 40 \\
HIP 2397B & 7.6033 & 22.7469 & 10.5; IR & 14.552$\pm$0.012 & 13.370$\pm$0.052 & -0.040$\pm$0.053 & HIP 2397 & K2 & 117.1 & 4063.4 & -0.220$\pm$0.09 & 1, 1, 1, 17, 41 \\
HD 3861B & 10.2945 & 9.3556 & 13.5; IR & 15.690$\pm$0.006 & 13.988$\pm$0.006 & 0.101$\pm$0.008 & HD 3861A & F8 & 16.5 & 554.2 & 0.137$\pm$0.01 & 42, 42, 2, 2, 25 \\
HIP 6407B & 20.5707 & 3.5232 & 11.0; IR & 15.345$\pm$0.005 & 14.203$\pm$0.006 & -0.121$\pm$0.008 & HIP 6407 / NLTT 4558 & G5 & 44.9 & 2105.8 & -0.020$\pm$0.03 & 1, 1, 2, 2, 25 \\
HIP 9269B & 29.7950 & 33.2087 & 16.0; IR & 16.126$\pm$0.017 & 14.298$\pm$0.019 & 0.165$\pm$0.025 & HIP 9269 & G9 & 52.1 & 1290.3 & 0.280$\pm$0.06 & 1, 1, 1, 1, 25 \\
SDSS J020735.60+135556.3 & 31.8983 & 13.9323 & 13.0; opt & 15.369$\pm$0.005 & 13.829$\pm$0.005 & 0.047$\pm$0.007 & G 3-40 & M2 & 72.5 & 2697.0 & 0.118$\pm$0.01 & 43, 43, 2, 2, 10 \\
GJ 1048B & 38.9997 & -23.5223 & 11.0; IR & 13.660$\pm$0.500 & 12.170$\pm$0.080 & 0.227$\pm$0.506 & GJ 1048A & K3.5 & 11.9 & 256.4 & 0.070$\pm$0.09 & 44, 44, 45, 45, 25 \\
HIP 26653B & 84.9564 & 52.8999 & 11.5; IR & 14.670$\pm$0.020 & 13.298$\pm$0.029 & 0.040$\pm$0.035 & HIP 26653 & G5 & 27.0 & 747.9 & -0.010$\pm$0.12 & 1, 1, 46, 17, 25 \\
LSPM J0632+5053B & 98.2021 & 50.8931 & 11.5; IR & 16.612$\pm$0.027 & 15.172$\pm$0.058 & 0.108$\pm$0.064 & LSPM J0632+5053 & & 47.4 & 4029.0 & 0.050 & 1, 1, 1, 17, 47 \\
HD 46588B & 101.6120 & 79.5835 & 19.0; IR & 16.097$\pm$0.093 & 14.600$\pm$0.090 & -0.130$\pm$0.129 & HD 46588A & F7 & 79.2 & 1441.4 & -0.085$\pm$0.13 & 48, 48, 5, 5, 25 \\
2MASS J07481084-2731309 & 117.0448 & -27.5251 & [10.0] & 16.024$\pm$0.098 & 14.945$\pm$0.149 & -0.109$\pm$0.178 & HD 63487 & G2  & 72.2 & 5224.0 & 0.075 & 6, 0, 5, 5, 49 \\
eta Cnc B & 128.1325 & 20.4502 & 13.0; IR & 17.737$\pm$0.035 & 16.523$\pm$0.036 & -0.279$\pm$0.050 & eta Cnc A & K3 & 164.0 & 15530.8 & 0.145$\pm$0.04 & 50, 50, 2, 2, 25 \\
SDSS J085836.98+271050.8 & 134.6539 & 27.1810 & 10.0; opt & 15.011$\pm$0.003 & 13.594$\pm$0.004 & 0.229$\pm$0.005 & LP 312-49 / NLTT 20640 & M4 & 15.4 & 816.2 & -0.133$\pm$0.02 & 51, 51, 2, 2, 10 \\
HD 89744B & 155.5620 & 41.2407 & 10.0; opt & 14.850$\pm$0.040 & 13.580$\pm$0.040 & 0.082$\pm$0.057 & HD 89744A & F7 & 63.1 & 2429.3 & 0.235$\pm$0.09 & 52, 52, 53, 53, 25 \\
Gl 417B & 168.1068 & 35.8035 & 14.5; opt & 14.470$\pm$0.020 & 12.670$\pm$0.030 & 0.222$\pm$0.036 & Gl 417A & G2 & 90.0 & 2097.0 & 0.115$\pm$0.03 & 54, 54, 46, 45, 25 \\
NLTT 27966B & 174.1654 & 48.8778 & 14.0; IR & 16.106$\pm$0.019 & 14.669$\pm$0.054 & -0.060$\pm$0.057 & NLTT 27966 & & 15.9 & 833.3 & -0.440$\pm$0.03 & 1, 1, 1, 17, 25 \\
HIP 59933B & 184.4020 & 14.4534 & 11.0; IR & 15.995$\pm$0.008 & 14.620$\pm$0.010 & 0.112$\pm$0.013 & HIP 59933 & F8 & 38.1 & 2560.3 & 0.140$\pm$0.04 & 1, 1, 2, 2, 41 \\
LSPM J1336+2541B & 204.1043 & 25.6769 & 14.0; IR & 16.935$\pm$0.017 & 15.478$\pm$0.016 & -0.040$\pm$0.023 & LSPM J1336+2541 & & 121.7 & 9582.7 & -0.483$\pm$0.02 & 1, 1, 2, 2, 10 \\
2MASS J13475911-7610054 & 206.9960 & -76.1681 & 10.0; opt & 13.671$\pm$0.033 & 12.554$\pm$0.032 & -0.071$\pm$0.046 & 2MASS J13475680-7610199 & M1 & 16.8 & 525.8 & -0.478$\pm$0.08 & 55, 56, 5, 5, 3 \\
PM I13518+4157B & 207.9478 & 41.9633 & 11.5; IR & 15.018$\pm$0.032 & 13.837$\pm$0.040 & -0.151$\pm$0.051 & PM I13518+4157 & & 21.6 & 1013.0 & 0.061$\pm$0.01 & 1, 1, 17, 17, 25 \\
G 239-25B & 220.5919 & 66.0555 & 10.0; IR & 11.440$\pm$0.030 & 10.300$\pm$0.070 & -0.048$\pm$0.076 & G 239-25A & M3 & 2.9 & 31.7 & -0.240$\pm$0.04 & 57, 58, 45, 45, 25 \\
HD 130948B & 222.5667 & 23.9116 & 14.0; IR & 13.200$\pm$0.080 & 11.690$\pm$0.040 & 0.013$\pm$0.089 & HD 130948A & F9 & 2.6 & 48.1 & -0.010$\pm$0.12 & 59, 60, 61, 61, 25 \\
HIP 73169B & 224.2985 & -6.3240 & 12.5; IR & 15.929$\pm$0.020 & 14.446$\pm$0.016 & 0.027$\pm$0.026 & HIP 73169 & M0 & 29.1 & 1307.8 & -0.010 & 1, 1, 1, 1, 15 \\
beta Cir B & 229.3400 & -58.8583 & 11.0; IR & 14.408$\pm$0.004 & 13.196$\pm$0.043 & -0.051$\pm$0.043 & beta Cir & A3 & 216.0 & 6199.2 & 0.200 & 62, 62, 63, 5, 64 \\
Gl 584C & 230.8443 & 30.2489 & 18.0; opt & 16.052$\pm$0.007 & 14.376$\pm$0.007 & -0.061$\pm$0.010 & Gl 584AB & G2$+$G2 & 194.0 & 3464.8 & -0.010$\pm$0.04 & 54, 54, 2, 2, 25 \\
GJ 618.1B & 245.1090 & -4.2755 & 12.5; opt & 15.180$\pm$0.050 & 13.550$\pm$0.040 & 0.174$\pm$0.064 & GJ 618.1A & K7 & 35.9 & 1166.7 & 0.370$\pm$0.09 & 52, 52, 45, 45, 31 \\
PSO J259.4554+29.9198 & 259.4554 & 29.9199 & [13.5] & 15.875$\pm$0.014 & 14.488$\pm$0.022 & -0.215$\pm$0.026 & 2MASS J17174844+2955095 & & 11.6 & 710.1 & 0.019$\pm$0.01 & 6, 0, 26, 27, 10 \\
LTT 7251B & 273.9548 & -23.8126 & 17.0; IR & 16.930$\pm$0.090 & 15.445$\pm$0.050 & -0.280$\pm$0.103 & LTT 7251 & G8 & 14.7 & 554.8 & -0.150$\pm$0.04 & 65, 65, 65, 65, 25 \\
zeta Del B & 308.8294 & 14.6710 & 15.0; IR & 17.180$\pm$0.090 & 15.400$\pm$0.070 & 0.140$\pm$0.114 & zeta Del & A3 & 13.5 & 903.1 & -0.050 & 66, 66, 66, 66, 64 \\
SDSS J213154.43-011939.3 & 322.9770 & -1.3273 & 19.0; opt & 17.230$\pm$0.015 & 15.545$\pm$0.210 & 0.058$\pm$0.211 & NLTT 51469AB & M3$+$M6 & 83.0 & 3859.5 & -0.411$\pm$0.02 & 67, 68, 68, 17, 10 \\
HD 221356D & 352.8791 & -4.0897 & 11.0; IR & 13.763$\pm$0.038 & 12.755$\pm$0.025 & -0.255$\pm$0.045 & HD 221356ABC & F7$+$M8$+$L3 & 12.1 & 315.0 & -0.190$\pm$0.06 & 69, 69, 69, 69, 25 \\
\enddata

\tablecomments{L dwarfs spectral types starts at L0 = 10.0, continuing after the M spectral types. The reference format follows Table \ref{tab:Mtab}. The indexing of the references is the same as Table \ref{tab:Mtab} for repeated references. The references not listed under Table \ref{tab:Mtab} are listed below. \\
References: 37 = \citet{Kirk01a}, 38 = \citet{Legg10a}, 39 = \citet{Eros99c}, 40 = \citet{lamost}, 41 = \citet{Kiefer2019}, 42 = \citet{Scho16}, 43 = \citet{Hawl02}, 44 = \citet{Gizi01a}, 45 = \citet{Fahe12}, 46 = \citet{Best20a}, 47 = \citet{Bochanski2018}, 48 = \citet{Lout11}, 49 = \citet{Casali2020}, 50 = \citet{Zhan10}, 51 = \citet{West08a}, 52 = \citet{Wils01b}, 53 = \citet{Dupuy2012}, 54 = \citet{Kirk00}, 55 = \citet{Kend07a}, 56 = \citet{Phan08a}, 57 = \citet{Goli04b}, 58 = \citet{Forv04}, 59 = \citet{Pott02}, 60 = \citet{Goto02}, 61 = \citet{Dupu09a}, 62 = \citet{Smit15}, 63 = \citet{Minn17}, 64 = \citet{Erspamer2003}, 65 = \citet{Smit18}, 66 = \citet{DeRo14b}, 67 = \citet{Chiu06}, 68 = \citet{Gauz19}, 69 = \citet{Gauz12}.
}

\end{deluxetable}
\end{longrotatetable}

\section*{acknowledgments}
This work has benefited from The UltracoolSheet at http://bit.ly/UltracoolSheet, maintained by Will Best, Trent Dupuy, Michael Liu, Rob Siverd, and Zhoujian Zhang, and developed from compilations by \citet{Dupuy2012}, \citet{Dupuy2013}, \citet{Liu2016}, \citet{Best2018}, and \citet{Best2020b}. We would like to give a special thanks to Eugene Magnier, Robert Siverd, Jamie Tayar, Zachary Claytor, and Ryan Dungee for helpful discussions regarding data analysis. The Pan-STARRS1 Surveys (PS1) and the PS1 public science archive have been made possible through contributions by the Institute for Astronomy, the University of Hawaii, the Pan-STARRS Project Office, the Max-Planck Society and its participating institutes, the Max Planck Institute for Astronomy, Heidelberg and the Max Planck Institute for Extraterrestrial Physics, Garching, The Johns Hopkins University, Durham University, the University of Edinburgh, the Queen's University Belfast, the Harvard-Smithsonian Center for Astrophysics, the Las Cumbres Observatory Global Telescope Network Incorporated, the National Central University of Taiwan, the Space Telescope Science Institute, the National Aeronautics and Space Administration under Grant No. NNX08AR22G issued through the Planetary Science Division of the NASA Science Mission Directorate, the National Science Foundation Grant No. AST-1238877, the University of Maryland, Eotvos Lorand University (ELTE), the Los Alamos National Laboratory, and the Gordon and Betty Moore Foundation. This publication makes use of data products from the Two Micron All Sky Survey, which is a joint project of the University of Massachusetts and the Infrared Processing and Analysis Center/California Institute of Technology, funded by the National Aeronautics and Space Administration and the National Science Foundation. This research has made use of the SIMBAD database and NASA/IPAC Infrared Science Archive. The UKIDSS project is defined in \citealt{UKIDSS2007}. UKIDSS uses the UKIRT Wide Field Camera (WFCAM; \citealt{Casali2007}) and a photometric system described in \citealt{Hewett2006}. The pipeline processing and science archive are described in \citet{Irwin2004} and \citet{Hambly2008}. The UHS is a partnership between the UK STFC, The University of Hawaii, The University of Arizona, Lockheed Martin and NASA. We have used data from the J-band data release described in detail in \citet{UHS_J2018} and the K-band data release described in \citet{UHS_K2018}. Some data in this work were obtained on Maunakea, the authors wish to recognize and acknowledge the very significant cultural role and reverence that the summit of Maunakea has always had within the indigenous Hawaiian community. We are most fortunate to have the opportunity to conduct observations from this mountain. 

\bibliography{bibFile}{}
\bibliographystyle{aasjournal}

\end{CJK*}
\end{document}